# FLRW cosmology with EDSFD parametrization

J. K. Singh[a], Ritika Nagpal[b]

Department of Mathematics, Netaji Subhas University of Technology, New Delhi 110078, India

**Abstract** In this paper, we study a cosmological model in the background of Friedmann–Lemaitre–Robertson–Walker (FLRW) space time by assuming an appropriate parametrization in the form of a differential equation in terms of energy density of scalar field $\rho_\phi$, which is defined as Energy Density Scalar Field Differential equation (EDSFD) parametrization. The EDSFD parametrization leads to a required phase transition from early deceleration to present cosmic acceleration. This parametrization is used to reconstruct the equation of state parameter $\omega_\phi$ in terms of redshift $z$ i.e. $\omega_\phi(z)$ to examine the evolutionary history of the universe in a flat FLRW space time. Here, we constrain the model parameter using the various observational datasets of Hubble parameter $H(z)$, latest Union 2.1 compilation dataset $SNeIa$, $BAO$, joint dataset $H(z)+SNeIa$ and $H(z)+SNeIa+BAO$ for detail analysis of the behavior of physical parameters and we find its best fit present value. Also, we discuss the dynamics of reheating phase after inflation, analyse the behaviors of the physical features using some diagnostic tools, and examine the viability of our parametric model.

## 1 Introduction

Since 1998, a substantial amount of observational data coming from various probes such as Type Ia Supernovae ($SNeIa$) [1], Cosmic Microwave Radiation Background (CMB) [2], Large Scale Structures (LSS) [3,4], Wilkinson Microwave Anisotropy Probe (WMAP) [5,6], Sloan Digital Sky Survey (SDSS) [7], Baryon Acoustic Oscillations (BAO) [8], weak lensing [9] and recent Planck collaboration [10,11] has shown and confirmed the accelerating expansion of the Universe. Figuring out the reason of acceleration of the Universe has been a major challenge to our cosmologists. To deal with this problem, there could be two possible actions: the first approach is to modify the energy momentum tensor (EMT) which likely contribute some amount of anti gravitational force to produce acceleration in the Universe, another way is to modify the geometry part in the Einstein Hilbert (EH) action, which is same as to modify General Theory of Relativity (GR).

In order to keep GR unaltered and yield an acceleration, a negative pressure fluid must be added to the content of the Universe. In this structure, Einstein suggested the cosmological constant $\Lambda$, which comes into sight as the most fascinating candidate to the dark energy (DE) because it acts on the field equations like a fluid with $p_\Lambda = -\rho_\Lambda$ and it can be related with the point of zero energy in the quantum fields. However, the $\Lambda$ leads to a huge variation between observation and theory despite its constancy with numbers of the cosmological data [16]. This variation has generated DE models beyond the $\Lambda$. Such DE models construct some unspecified symmetries which reduce the effect of the vacuum energy (VE) contribution. If the vacuum energy (VE) cannot be dropped off, then an additional attempt to reduce the inconsistency between theory and observation is to assume that $\Lambda$ develop gradually with time. This expectation signify that the dark matter (DM) and the vacuum energy (VE) are not conserved independently. The phenomenological DE models specified by $\omega = \frac{p_{DE}}{\rho_{DE}}$, the ratio of pressure to density, and the vacuum decay scenarios are built to get the standard $\Lambda CDM$ model with equation of state (EoS) $\omega_\Lambda = -1$ as a special case. This is due the success of the $\Lambda$-term in describing the current observations. Several authors have discussed the DE models using various recent observations [12–15].

The issues related to the fine tuning problem and cosmic coincidence problem, which are connected with standard $\Lambda CDM$ motivate us to work on some other types of DE models [16,17]. The DE phenomenon as an effect of dissipation like bulk viscosity investigated in cosmic medium are available in many articles in the literature [18,19]. Therefore, the dynamical dark energy models came in to existence to the resolve the difficulties related $\Lambda CDM$ model [20,21].

[a] e-mail: jksingh@nsut.ac.in
[b] e-mail: ritikanagpal.math@gmail.com (corresponding author)

A large variety of scalar field DE models have been proposed in literature [22–25]. The most acceptable and popular DE is quintessence scalar field, which serves as a dynamical quantity with variable density in space time and can be attractive or repulsive both depending on the ratio of its kinetic energy (KE) and potential energy (PE). The quintessence scalar field can predict accelerated expanding universe provided its potential is chosen suitably so that the scalar field can roll slowly over there. In other words, if the KE of the scalar field is very less as compared to PE i.e. $\dot{\phi}^2 << V(\phi)$, acceleration in the Universe can be anticipated [26–29].

Some of the authors have studied the history of cosmic evolution of the Universe using *reconstruction method* [30,31]. Reconstruction technique in the framework of DE was first studied by Starobinsky [32]. The reconstruction method are of two types: (i) parametric and (ii) non parametric. In the parametric method, the construction of a cosmological model is based on a specific parameterization of the various cosmological parameters. The cosmic dynamics can be discussed by constraining the model parameters involved in the parametrization using various observational datasets. However, this technique has some limitations about the exact nature of DE due to the assumption of a cosmic parameter [33–35]. In this method, some models seems diverging sometimes to describe the future cosmic scenario and some models are consistent only at very low redshift ($z << 1$). In this aspect, a non parametrizing method has some preference over parametrization method because of the direct understanding of the cosmic scenario through observational data. However, this approach has also some drawbacks [36–38]. Therefore, we do not have a well motivated theory which could explain all the phenomenon of the Universe at present.

It has been seen that parametrization method can better explain the transition history of the Universe successfully from early deceleration to late times acceleration in many of the models. So it is feasible to endorse a parametrization approach which can explain the future cosmological models with better efficiency and motivates us to construct the model of the Universe with possible phase transition [39–41]. Cunha et al. [39] studied the transition redshift by expanding the deceleration parameter by using two parameterizations: $q = q_0 + q_1 z$ and $q = q_0 + q_1 z (1 + z)^{-1}$, where the parameters $q_0, q_1$ are constrained by the supernovae data. Akarsu et al. [40] probed the kinematics and fate of the Universe with linearly time varying deceleration parameter $q = q_0 + q_1(1 - t/t_0)$. Xu et al. [41] discussed the FLRW cosmological model using a parametrized deceleration parameter $q(a) = q_0 + q_1(1 - a)$, where the parameters $q_0, q_1$ are constrained by the observational data $SNeIa$ and $CMB$. Recently, many cosmological models have been discovered to explain the dynamics of the Universe in $f(R, T)$ gravity using parametrization approach of different cosmological parameters [42–45]. Singh et al. [42] reviewed the FLRW model using a specific form of parametrization on Hubble parameter $H(t) = \alpha \, t \, ln\left(\frac{c - \lambda_2 \tan^{-1} t}{t}\right)$, where $\alpha, \lambda_2, c$ are arbitrary constants which produces the bouncing scenario in $f(R, T)$ gravity. Singh et al. [43] constructed a modified Chaplygin gas cosmological model in f(R,T) gravity with particle creation using a specific parametrization on Hubble parameter $H(t) = \frac{\beta t^{k_1}}{(t^{k_2} + \alpha)^{k_3}}$, where $\alpha, \beta \neq 0$, $k_1, k_2, k_3$ are real constants. The FLRW cosmological models with quark and strange quark matters in f(R,T) gravity is discussed by using two parametrization: (i) the $q = \frac{-\alpha}{t^2} + (\beta - 1)$, where $\alpha > 0, \beta > 1$ are model parameter which are responsible for the dynamics of the Universe, (ii) $a(t) = e^{k_1 t} t^{k_2}$, where the parameters $k_1$ and $k_2$ are non-negative and dimensionless [44]. An exact cosmological solution of Einstein field equations (EFE) is derived for a dynamical vacuum energy in $f(R, T)$ gravity in the background of a homogeneous and isotropic FLRW space time by considering an appropriate parametrization of the Hubble parameter $H(a) = \alpha(1 + a^{-n})$, where $\alpha > 0$ and $n > 1$ are the model parameters, which are constrained through recent observational data [45].

In literature, there are many parametrized forms proposed by many authors in the form of EoS parameter $\omega$ to make the model consistent with observations [46–48]. Linder [46] explored the the expansion history of the Universe using a new parametrization of the EoS parameter $\omega(a) = \omega_0 + \omega_a(1 - a)$, where where $\omega_0$ and $\omega_a$ are model parameters. A hyperbolic solution for a flat FLRW model of the universe in $f(R, T)$ gravity is discussed by adopting a parametrization of scale factor $a(t) = \sinh^{\frac{1}{n}}(\beta t)$, where $\beta$ and $n > 0$ are arbitrary constants [47]. Nojiri et al. [48] studied the modified gravity to unify inflation with dark energy and dark matter by assuming the Hubble parameter of the form $H(t) = c_1(-t + t_f)^\alpha + c_2(-t + t_s)^\beta + c_1 t^\alpha + g$, where $c_1, c_2, \alpha, \beta, g$ are constants. However, the parametrization of any specific cosmological parameter plays a significant role to study the cosmic scenario but it has been observed that the parametrization of the energy density of DE contributes more compacted constraints than any other cosmological parameter [49,50]. Therefore in this paper, we would like to stress more on the form of the energy density of dark energy which yields a phase transition in the Universe and attempt to explain the cosmic phenomenon successfully using some observational datasets. Motivated by the above parametrizations in which many of them are linear, we have adopted the parametric reconstruction approach to study the DE model in GR at present study. In this paper, we study a FLRW cosmological model by assuming an appropriate parametrization in the form of a differential equation in terms of energy density of scalar field $\rho_\phi$, which is defined as Energy Density Scalar Field Differential equation (EDSFD) parametrization and is unique of its own kind. The EDSFD parametrization leads to a required phase transition from early deceleration to present cosmic acceleration and is capable to explain several stages of evolution of the Universe and form the model which is consistent with current observational data.

The advancement of constraints on inflation followed by reheating with well planned approaches in observational cosmology will be an essential part of research in near future. If we restrict the range of the observationally enabled models of inflation, It is difficult to understand the uncertainties in the predictions of the spectral index. The study on reheating have also shown the feasibility of production of remains like solitons and cosmic defects which help us to constrain the variety of scenarios [51]. The preheating stage can contain extremely non perturbative procedures, during which the Universe gets populated through the parametric resonances. Such resonances appear as the inflaton condensate which starts oscillation about its minimum potential soon after inflation. Theses oscillations induce a plausible time dependent couplings between the inflaton and the matter. The non linear dynamics can conduct to a lot of interesting facts, e.g. the creation of solitons that can slow down the thermalisation process which is essential as an initial condition for Big Bang nucleosynthesis (BBN), field structures yielding self similarly in a turbulent manner [52,53], non thermal phase transitions and the formation of cosmic defects [54–56].

There are three important research areas of reheating: (i) the theoretical part which consider the more realistic high energy physics models [57–59] involving fermions and gauge bosons along with the scalars, in the search for a unified illustration of our cosmic history, (ii) the direction for future work regarding the happening of various phases of reheating: from the non perturbative particle creation during the preheating phase followed by non linear classical evolution, to the late time approach to a radiation dominated era of expansion of the Universe in the thermal equilibrium [60,61], (iii) a lot of work has been devoted to the reheating constraints to inflationary models [62–64]. Kofman et al. [65] explored the reheating process of the Universe in various inflationary models in which they found that the decay of the inflaton field starts with a phase of explosive creation of particles at a stage of a broad parametric resonance and later finished together. Kofman et al. [66] studied the contemporary growth of the Reheating theory. He stated that all the matter constituents of the Universe had been produced during the reheating process after inflation. Kofman et al. [67] examined that the reheating after inflation takes place due to the creation of particle by the oscillating inflaton field. He also analysed the perturbative approach to reheating as well as the effects beyond the perturbation theory. Bassett et al. [68] also reviewed the theory of inflation with single and multiple fields with certain observation on the dynamics of adiabatic along with entropy and isocurvature perturbations which provides the basic probe of inflationary models.

The work of this paper has been organized as follows: Sect. 1 provides a brief formal presentation on dark energy and brief proposal about the late time cosmic acceleration which have been confirmed by various probes. In Sect. 2, we consider the Einstien field equations and obtain its GR solutions using a parametrization consists of specific form of scalar field differential equation. In Sect. 3, we constrain the model parameter for the detailed analysis on the behaviour of physical parameters using the observational datasets of Hubble parameter $H(z)$, latest union 2.1 compilation datasets $SNeIa$, $BAO$, joint datasets $H(z)+SNeIa$ and $H(z)+SNeIa+BAO$, and we find the best fit present values of the cosmological parameters. We study the dynamics of our parametric model and briefly analyse the behaviours of the physical features using statistical analysis of cosmological parameters of the model in Sect. 4. Also, we explain the viability of our model via energy conditions in Sect. 5. We discuss the dynamics of reheating stage after inflation of our parametric model in Sect. 6. Finally, we review the consistency of our obtained model using data analysis and conclude the outcomes of the model in Sect. 7.

## 2 Einstien field equations and its solutions

The Einstien field equations (EFE) in the background of a flat FLRW space time

$$ds^2 = dt^2 - a^2(t)(dx^2 + dy^2 + dz^2), \tag{1}$$

where $a(t)$ being the scale factor of the Universe are given by

$$\left(\frac{\dot{a}}{a}\right)^2 = \frac{1}{3}(\rho_{total}) = \frac{1}{3}(\rho_m + \rho_\phi), \tag{2}$$

$$\left(\frac{\dot{a}}{a}\right)^2 + 2\left(\frac{\ddot{a}}{a}\right) = -p_{total} = -p_\phi, \tag{3}$$

in the units of $8\pi G = c = 1$, where an overhead dot signifies the time derivatives. The energy density of matter, energy density of scalar field and pressure of scalar field are represented by $\rho_m$, $\rho_\phi$ and $p_\phi$ respectively. Here, we assume that the total amount of energy $\rho_{total}$ and matter source $p_{total}$ contained in the Universe are composed of two types of fluid, especially the one corresponds to non relativistic matter or the pressure less cold matter and the other scalar field $\phi$, which serves as a candidate of dynamical DE i.e. $\phi$ varies with time $t$ and describes the early inflation as well as the late time cosmic acceleration.

The following action defines the most generic models of inflation which is consistent with observations [69] in which a single scalar field $\phi$ is termed as the inflaton sources the accelerated expansion of the universe as

$$S = \frac{c^4}{16\pi G} \int \left(-R\sqrt{-g}\right) d^4x + S_\phi + S_m, \quad (4)$$

where $S_\phi$ is the action for the scalar field. Here, the model is minimally coupled to gravity with canonical kinetic terms. $S_m$ is the matter action term which includes the entire information related the other components of the matter, containing the Standard Model Lagrangian (SML) and the terms explaining the couplings of the inflaton to other fields [51]. We normalize Eq. (4) by taking $8\pi G = c = 1$.

The action of the scalar field is given by

$$S_\phi = \int \left[\frac{1}{2}\partial_\mu\phi\partial^\mu\phi - V(\phi)\right]\sqrt{-g}d^4x. \quad (5)$$

In order to explain the isotropy and homogeneity of the model, we need the dominant component of the scalar field $\phi$ which varies with $t$ only. This scalar condensation caused the classical background configuration during the inflation and the initial stages of reheating. As the scalar field $\phi$ depends on time, therefore it may be assumed as perfect fluid with energy density $\rho_\phi$ and pressure $p_\phi$. When we suppose that the scalar field $\phi$ is the only source of DE having potential $V(\phi)$ which interacts with itself, so we can take energy density $\rho_\phi$ and pressure $p_\phi$ as the canonical components of scalar field $\phi$ in the framework of FLRW cosmology given by [47,70,71]

$$\rho_\phi = \frac{\dot\phi^2}{2} + V(\phi), \quad (6)$$

$$p_\phi = \frac{\dot\phi^2}{2} - V(\phi), \quad (7)$$

where the potential energy $V(\phi)$ and kinetic energy $\frac{\dot\phi^2}{2}$ are function of scalar field $\phi$ correspond to each pair of $(t, x)$ in space time.

The conservation of energy momentum tensor in GR read as

$$\dot\rho_m + 3\left(\frac{\dot a}{a}\right)\rho_m = 0, \quad (8)$$

$$\dot\rho_\phi + 3\left(\frac{\dot a}{a}\right)(\rho_\phi + p_\phi) = 0. \quad (9)$$

Eq. (8) yields

$$\rho_m = \rho_{m_0} a^{-3}, \quad (10)$$

where $\rho_{m_0}$ is an arbitrary integration constant and evaluated as the present value of the energy density of the matter. Solving Eq. (9), we get

$$\dot\rho_\phi = -3\left(\frac{\dot a}{a}\right)(1 + \omega_\phi)\rho_\phi, \quad (11)$$

where $\omega_\phi = \frac{p_\phi}{\rho_\phi}$ is the EoS parameter of the scalar field $\phi$. Using Eq. (9), the EoS parameter is given by

$$\omega_\phi = -1 - \frac{1}{3}a\frac{\rho_\phi'}{\rho_\phi}, \quad (12)$$

where prime denotes the derivative w.r.t. scale factor $a$. In order to solve a system of Eqs. (2), (3), (8) and (9) which contains only three independent equations with four unknowns $a$, $\rho_m$, $\rho_\phi$ and $p_\phi$, we need some additional constraint equations.

Here we assume an appropriate parametrization as a differential equation in $\rho_\phi$ of the form

$$\rho_\phi' + \alpha f(a)\rho_\phi = 0, \quad (13)$$

where $f(a) = \alpha^2\left[1 + \frac{1}{a\sqrt{1+a^{2\alpha}}sinh^{-1}\left(\frac{1}{a}\right)^\alpha}\right]$.

The function $f(a)$ has been taken in such way that the term $sinh^{-1}\left(\frac{1}{a}\right)$ contained in $f(a)$ can produce a signature flipping behaviour. Therefore, the Eq. (13) shows concave downward behavior i.e. having decreasing slope at each point with respect to time. The transition phase of deceleration parameter $q$ from positive to negative exhibits cosmic acceleration at present and late times onwards along with early deceleration with possible structures formation in the Universe. These types of various parametrization in different mathematical forms of cosmological parameters have already been used in cosmology to study dark energy models [39,42,46] because they neither presume the validity of any gravitational theory nor effect the model by violating any physical and geometrical properties [72].

The general solution of the above differential equation (13) is given by

$$\rho_\phi = e^{-\alpha a} sinh^{-1}\left(\frac{1}{a}\right)^\alpha, \quad (14)$$

where $\alpha \in (0, 1)$ is the model parameter.

Using the relation of redshift $z$ and scale factor $a$ given by $\frac{a}{a_0} = \frac{1}{(1+z)}$, where $a_0 = 1$ is the present value of scale factor, the energy density of scalar field $\rho_\phi$ in terms of redshift reads as

$$\rho_\phi(z) = e^{\frac{-\alpha}{1+z}} sinh^{-1}(1+z)^\alpha, \quad (15)$$

and

$$\rho_{\phi_0} = e^{-\alpha} sinh^{-1}(1). \quad (16)$$

Equations (15) and (16) yield

$$\rho_\phi(z) = \frac{\rho_{\phi_0}}{sinh^{-1}(1)} e^{\frac{\alpha z}{1+z}} sinh^{-1}(1+z)^\alpha, \quad (17)$$

where $\rho_{\phi_0}$ is the current value of the energy density of scalar field. Also using Eq. (10), energy density of matter field $\rho_m$ can be written in terms of redshift $z$ as

$$\rho_m(z) = \rho_{m_0}(1+z)^3 \quad (18)$$

Combining Eqs. (2), (17) and (18), we have

$$3H^2 = \rho_{m_0}(1+z)^3 + \frac{\rho_{\phi_0}}{sinh^{-1}(1)} e^{\frac{\alpha z}{1+z}} sinh^{-1}(1+z)^\alpha, \quad (19)$$

Now, we consider the density parameter $\Omega = \frac{\rho}{\rho_c}$ which describe the whole content of the Universe, where $\rho_c = \frac{3H^2}{(8\pi G)^2}$ is the critical density of the Universe. In the normalized unit, $8\pi G = 1$.

Equation (19) in terms of density parameter of matter and scalar field can be expressed as

$$H^2 = H_0^2 \left[ \Omega_{m_0}(1+z)^3 + \frac{\Omega_{\phi_0}}{sinh^{-1}(1)} e^{\frac{\alpha z}{1+z}} sinh^{-1}(1+z)^\alpha \right], \quad (20)$$

where $\Omega_{m_0} = \frac{\rho_{m_0}}{3H_0^2}$ and $\Omega_{\phi_0} = \frac{\rho_{\phi_0}}{3H_0^2}$ are the present values of matter and scalar field density parameters respectively.

As it is defined a dimensionless quantity $q$ which measures the acceleration in the Universe and is known as deceleration parameter (DP). Obviously $q < 0$ affirms that the Universe undergoes through an accelerated expanding phase whereas $q > 0$ signifies that expansion in the Universe is decelerating. The DP $q$ in terms of $a$ and the Hubble parameter $H$ is given by

$$q = -\frac{\ddot{a}a}{\dot{a}^2} = -1 - \frac{\dot{H}}{H^2}. \quad (21)$$

In our model, using Eq. (20), the expression for DP in terms of $a$ and $z$ are given by

$$q(a) = -1 - \frac{a\left(\frac{-3\Omega_{m_0}}{a^4} - \frac{\alpha e^{\alpha(1-a)} cosech^{-1}(a)^{\alpha-1}\left[\sqrt{1+\frac{1}{a^2}}+(1+a^2)cosech(a)\right]\Omega_{\phi_0}}{(1+a^2)sinh^{-1}(1)}\right)}{2\left(\frac{\Omega_{m_0}}{a^3} + \frac{e^{\alpha(1-a)} cosech^{-1}(a)^\alpha \Omega_{\phi_0}}{sinh^{-1}(1)}\right)}, \quad (22)$$

$$q(z) = -1 + \frac{1}{2(1+z)\left[(1+z)^3\Omega_{m_0} + \frac{e^{\frac{\alpha z}{1+z}} cosech^{-1}(\frac{1}{1+z})^\alpha \Omega_{\phi_0}}{sinh^{-1}}\right]}$$
$$\left[3(1+z)^4 \Omega_{m_0} + \left\{\alpha e^{\frac{\alpha z}{1+z}} cosech^{-1}\left(\frac{1}{1+z}\right)^{\alpha-1}\right\}\right.$$
$$\times \left\{\sqrt{1+(1+z)^2} + \left(1+\frac{1}{(1+z)^2}\right) cosech^{-1}\left(\frac{1}{1+z}\right)\right\}$$
$$\left.\times \left\{\frac{\Omega_{\phi_0}(1+z)^2}{1+(1+z)^2 sinh^{-1}(1)}\right\}\right]. \quad (23)$$

Using relation (12), the EoS parameter for scalar field $\omega_\phi$ can be expressed in terms of $z$ and $a$ as

$$\omega_\phi(z) = -1 + \frac{\alpha}{3(1+z)}$$
$$+ \frac{\alpha(1+z)^\alpha}{\sqrt{1+(1+z)^{2\alpha}} sinh^{-1}(1+z)^\alpha}, \quad (24)$$

and

$$\omega_\phi(a) = -1 + \frac{\alpha a}{3} + \frac{\frac{\alpha}{a^\alpha}}{3\sqrt{1+\frac{1}{a^{2\alpha}}} sinh^{-1}\left(\frac{1}{a}\right)^\alpha}, \quad (25)$$

respectively.

The functional form of the density parameters $\Omega_m$ and $\Omega_\phi$ for matter and scalar field in terms of redshift $z$ are given by

$$\Omega_m(z) = \frac{1}{1 + \frac{e^{\frac{\alpha z}{1+z}} sinh^{-1} \Omega_{\phi_0}(1+z)^\alpha}{sinh^{-1}(1)\Omega_{m_0}(1+z)^3}}, \quad (26)$$

and

$$\Omega_\phi(z) = \frac{1}{1 + \frac{\Omega_{m_0} sinh^{-1}(1) e^{\frac{-\alpha z}{1+z}}[(1+z)^3 sinh^{-1}(1+z)^{-\alpha}]}{\Omega_{\phi_0}}}, \quad (27)$$

respectively.

In the next section, we have constrained the model parameters using various observational data set to find the best fit present values of cosmological parameters for the detailed analysis on the behaviour of physical parameters.

## 3 Statistical analysis of the model parameters

In recent advancement of observational cosmology, we study the early evolution, structure formation, and the properties of DM and DE of the Universe employing cosmic mechanism and ray detectors. There are various observational datasets e.g. $SDSS$, which comes up with the map of the galaxy distribution and encode the current variations in the Universe, $CMBR$ performs as the authentication of the big bang theory, $QUASARS$ brings out the matter between observer and quasars, $BAO$ estimates the large scale structures in the Universe to interpret the DE more desirable, observations from $SNeIa$ are the devices for computing the cosmic distances known as standard candles.

In this section, we compare our model with the $\Lambda CDM$ model using error bar plots of Hubble dataset and $SNeIa$ Union 2.1 compilation dataset. We have also constrained the value of Hubble parameter $H_0$ and model parameter $\alpha$ included in our model by employing some observational datasets $H(z)$, $SNeIa$, $BAO$ and their joint datasets $H(z) + SNeIa$ and $H(z) + SNeIa + BAO$ using statistical analysis method. These constrained values of the model parameter $\alpha$ have been used to discuss the physical features of the our model.

In Fig. 1, the error bar plots of Hubble dataset and $SNeIa$ Union 2.1 compilation dataset represent that both panels are fitted well while correlating our model with $\Lambda CDM$ using observational dataset $H(z)$ and $SNeIa$ Union 2.1 compilation data respectively.

### 3.1 Observational Constraints from H(z) data

As it is well known that the expansion rate of the Universe can be expressed in terms of an important cosmological parameter termed as Hubble parameter $H$ gained high popularity at present. In the series of observations, Hubble parameter data is one such useful and simple tool to study cosmological models as it has a direct relation with the expansion history of the Universe. The observations from Hubble parameter help us to illuminate power on the dark sector of the Universe which precisely includes the problem of dark energy, dark matter and dark ages. To make $H$ observable, some physical quantities such as length, time and redshift $z$ are required. The Hubble parameter $H$ in terms of $z$ can be written as

$$H(z) = -\frac{1}{(1+z)}\frac{dz}{dt}. \tag{28}$$

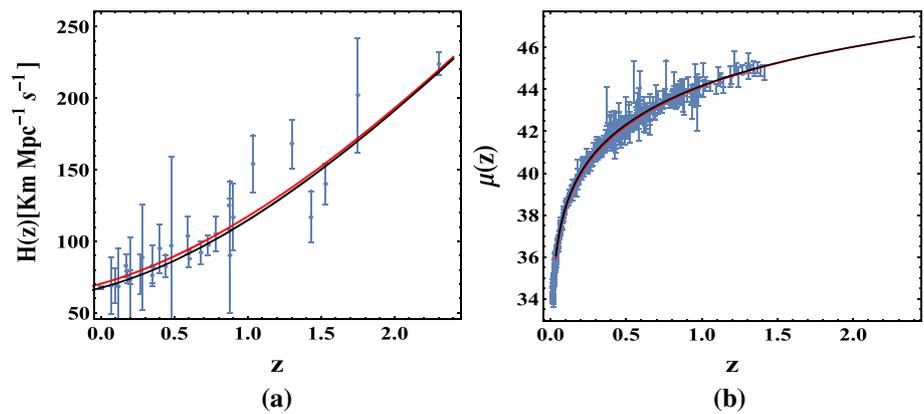

**Fig. 1** Figures (**a**) and (**b**) represent the comparison of our model and $\Lambda CDM$ model with blue color error bars of observational $H(z)$ 29 data points and $SNeIa$ Union 2.1 compilation data points respectively. Red lines indicate our model and black lines indicate the $\Lambda CDM$ model in both plots. We observe that our model fits well with both observational data points especially at $z < 1$

The constraints on the model parameter $\alpha$ can be obtained by minimizing the Chi-square value i.e. $\chi^2_{min}$, which is equivalent to the maximum likelihood analysis. The likelihood function $\chi^2_{OHD}(\alpha, H_0)$ can be computed as

$$\chi^2_{OHD}(\alpha, H_0) = \sum_{i=1}^{29}\left[\frac{H_{th}(\alpha, H_0, z_i) - H_{obs}(z_i)}{\sigma_{H(z_i)}}\right]^2, \tag{29}$$

where $H(z_i)$ is evaluated at redshift $z_i$, $OHD$ represents the observational Hubble data set, and $H_{th}$ and $H_{obs}$ represent the theoretical and observed value of Hubble parameter $H$ of our model respectively. The standard error in the observed value of $H$ is indicated by $\sigma_{H(z_i)}$.

### 3.2 Observational constraints from type ia supernova data

Here, we constrain the model parameter $\alpha$ by using the latest Union 2.1 compilation datasets $SNeIa$ from the observations of type Ia Supernova [73] and compare the results with $\Lambda CDM$. The Chi-square function $\chi^2_{OSN}(\alpha, H_0)$ can be expressed as

$$\chi^2_{OSN}(\alpha, H_0) = \sum_{i=1}^{580}\left[\frac{\mu_{th}(\alpha, H_0, z_i) - \mu_{obs}(z_i)}{\sigma_{\mu(z_i)}}\right]^2, \tag{30}$$

where $OSN$ is the observational $SNeIa$ dataset. $\mu_{th}$ and $\mu_{obs}$ are the theoretical and observed distance modulus of the model. The standard error in the observed value is denoted by $\sigma_{\mu(z_i)}$. Also, the distance modulus $\mu(z)$ is defined by

$$\mu(z) = m - M = 5 Log D_l(z) + \mu_0, \tag{31}$$

where $m$ and $M$ denote the apparent and absolute magnitudes of a standard candle, respectively. The luminosity distance $D_l(z)$ for flat Universe and the nuisance parameter $\mu_0$ are given by

$$D_l(z) = (1+z)H_0\int_0^z \frac{1}{H(z^*)}dz^*, \tag{32}$$

and

$$\mu_0 = 5 Log\left(\frac{H_0^{-1}}{Mpc}\right) + 25, \tag{33}$$

respectively.

We take the current value of density parameter for matter and scalar field ($\Omega_{m_0}$ and $\Omega_{\phi_0}$) as 0.27 and 0.73, respectively. Also, we use the present value of Hubble parameter from the most recent Planck 2018 results [10,11] as $H_0 = 67.4$ Km/s/Mpc to complete the dataset. The likelihood contours have been plotted in the $\alpha$-$H_0$ plane with $1\sigma$, $2\sigma$ and $3\sigma$ errors at the confidence levels 68.27%, 95.45% and 99.73%, respectively, around the best fit points by fitting our model with $H(z)$, $SNeIa$, $BAO$, $H(z)+SNeIa$ and $H(z)+SNeIa+BAO$ datasets respectively.

Here, the best fit values of the model parameters are $q_0 = -0.477$, $\alpha = 0.216746$ and $H_0 = 65.4940$ with $\chi^2_{OHD} = 18.3086$ confirmed by the likelihood contours in the $\alpha$-$H_0$ plane with $1\sigma$, $2\sigma$ and $3\sigma$ errors at the confidence levels 68.27%, 95.45% and 99.73% respectively for the Hubble data analysis around the best fit point (0.216746, 65.4940) shown in Fig. 2a. Similarly the best fit values of the model parameters are $q_0 = -0.540$, $\alpha = 0.128497$ and $H_0 = 66.4581$ with $\chi^2_{OHD} = 562.458$ which are confirmed by the likelihood contours in the $\alpha$-$H_0$ plane with $1\sigma$, $2\sigma$ and $3\sigma$ errors at the confidence levels 68.27%, 95.45% and 99.73% respectively for the $SNeIa$ data analysis around the best fit point (0.128497, 66.4581) shown in Fig. 2b.

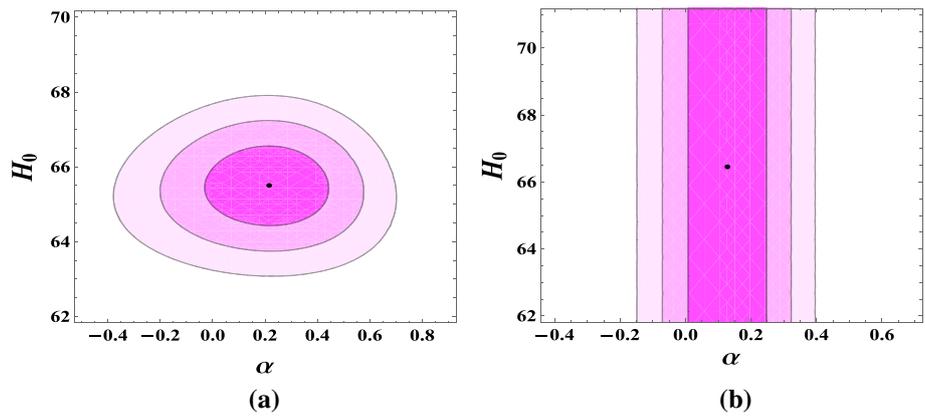

**Fig. 2** **a** and **b** Show the likelihood contours in the $\alpha$-$H_0$ plane for joint analysis $H(z) + SNeIa$ and $H(z) + SNeIa + BAO$, respectively. The dark shaded region shows the $1\sigma$ error, light shaded region shows the $2\sigma$ error and ultra light shaded region shows the $3\sigma$ error. Black dots represent the best fit values of the model parameter $\alpha$ and the values of $H_0$ in both plots

### 3.3 Observational constraints from baryon acoustic oscillations data

The $BAO$ includes computing the spatial distribution of galaxies to regulate the growing rate of cosmic structure within the comprehensive expansion of the universe. This contrast can differentiate between the various forms of DE theoretically. The patterns of galaxy clustering involve statistics about how cosmic structure is magnified from initial minor variations. This clustering encodes a powerful *average separation* between galaxies which could be used to construct the expansion history of the universe in a similar fashion to $SNeIa$ standard candles.

$$C^{-1} = \begin{bmatrix} 0.52552 & -0.03548 & -0.07733 & -0.00167 & -0.00532 & -0.00590 \\ -0.03548 & 24.97066 & -1.25461 & -0.02704 & -0.08633 & -0.09579 \\ -0.07733 & -1.25461 & 82.92948 & -0.05895 & -0.18819 & -0.20881 \\ -0.00167 & -0.02704 & -0.05895 & 2.91150 & -2.98873 & 1.43206 \\ -0.00532 & -0.08633 & -0.18819 & -2.98873 & 15.96834 & -7.70636 \\ -0.00590 & -0.09579 & -0.20881 & 1.43206 & -7.70636 & 15.28135 \end{bmatrix}$$

In this study, we examine a sample of $BAO$ distance measurements from various surveys, namely $SDSS(R)$ [74], 6dF Galaxy survey [75], $BOSSCMASS$ [76] and three parallel measurements from $WiggleZ$ survey [77].

The distance redshift ratio $d_z$ regarding $BAO$ measurements is given by

$$d_z = \frac{r_s(z^*)}{D_v(z)}, \tag{34}$$

where $z^*$ is the redshift at the time of photons decoupling which is taken as $z^* = 1090$ according as the Planck 2015 results [78]. The term $r_s(z^*)$ is co-moving sound horizon during decoupling of photons [79]. The dilation scale is defined as $D_v(z) = \left[\frac{d_A^2(z)z}{H(z)}\right]^{\frac{1}{3}}$, where $d_A(z)$ is the angular diameter distance.

The value of $\chi^2_{BAO}$ is given by [80]

$$\chi^2_{BAO} = A^T C^{-1} A, \tag{35}$$

where $A$ is a matrix given by

$$A = \begin{bmatrix} \frac{d_A(z^*)}{D_v(0.106)} - 30.84 \\ \frac{d_A(z^*)}{D_v(0.35)} - 10.33 \\ \frac{d_A(z^*)}{D_v(0.57)} - 6.72 \\ \frac{d_A(z^*)}{D_v(0.44)} - 8.41 \\ \frac{d_A(z^*)}{D_v(0.6)} - 6.66 \\ \frac{d_A(z^*)}{D_v(0.73)} - 5.43 \end{bmatrix}$$

and the inverse covariance matrix $C^{-1}$ [80] is approaching the correlation coefficients available in [81].

In $BAO$ data analysis, the good fit values of the model parameters are $q_0 = -0.626$, $\alpha = 0.007485$ and $H_0 = 65.0100$ with $\chi^2_{BAO} = 33.9581$, which are confirmed by the likelihood contours in the $\alpha$-$H_0$ plane with $1\sigma$, $2\sigma$ and $3\sigma$ errors at the confidence levels 68.27%, 95.45% and 99.73%, respectively around the best fit point (0.007485, 65.0100).

### 3.4 Joint tests: $H(z) + SNeIa$ data and $H(z) + SNeIa + BAO$ data

In order to obtain tighter constraints on the model parameters and to avoid the degeneracy in the observational datasets, we combine $H(z)$ and $SNeIa$ data, and $H(z)$, $SNeIa$ and $BAO$ data. Since $H(z)$, $SNeIa$ and $BAO$ datasets are obtained from the independent probes, the total likelihoods are considered to be the product of the likelihood of the probes $H(z)$

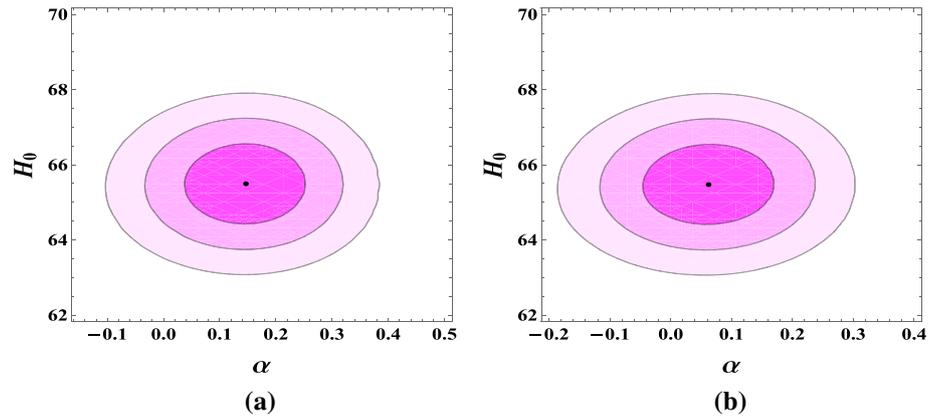

**Fig. 3** **a** and **b** Show the likelihood contours in the $\alpha$-$H_0$ plane for joint analysis $H(z) + SNeIa$ and $H(z) + SNeIa + BAO$, respectively. The dark shaded region shows the $1\sigma$ error, light shaded region shows the $2\sigma$ error and ultra light shaded region shows the $3\sigma$ error. Black dots represent the best fit values of the model parameter $\alpha$ and the values of $H_0$ in both plots

**Table 1** Summary of the numerical results of the model

| Dataset | $\chi^2_{min}$ | Hubble parameter $H_0$ (km/s/Mpc) | Model parameter $\alpha$ |
| --- | --- | --- | --- |
| $H(z)$ (29 points data ) | 18.3086 | 65.4940 | 0.216746 |
| $SNeIa$ (Union 2.1 compilation data) | 562.458 | 66.4581 | 0.128497 |
| $BAO$ | 33.9581 | 65.0100 | 0.007485 |
| $H(z) + SNeIa$ | 581.015 | 65.4934 | 0.145958 |
| $H(z) + SNeIa + BAO$ | 618.312 | 65.4800 | 0.062845 |

and $SNeIa$ and the product of the likelihood of the probes $H(z)$, $SNeIa$ and $BAO$. Therefore, we define

$$\chi^2_{total_{HS}} = \chi^2_{OHD} + \chi^2_{OSN}, \quad (36)$$

and

$$\chi^2_{total_{HSB}} = \chi^2_{OHD} + \chi^2_{OSN} + \chi^2_{BAO}, \quad (37)$$

respectively. In joint analysis $H(z) + SNeIa$, we obtained the best fit values of the model parameters are $q_0 = -0.527$, $\alpha = 0.145958$ and $H_0 = 65.4934$ with $\chi^2_{total_{HS}} = 581.015$ which are confirmed by the likelihood contours in the $\alpha$-$H_0$ plane with $1\sigma$, $2\sigma$ and $3\sigma$ errors at the confidence levels 68.27%, 95.45% and 99.73%, respectively around the best fit point (0.145958, 65.4934) shown in Fig. 3a. Similarly the best fit values of the model parameters are $q_0 = -0.586$, $\alpha = 0.062845$ and $H_0 = 65.4800$ with $\chi^2_{total_{HSB}} = 618.312$ which are confirmed by the likelihood contours in the $\alpha$-$H_0$ plane with $1\sigma$, $2\sigma$ and $3\sigma$ errors at the confidence levels 68.27%, 95.45% and 99.73%, respectively for joint analysis $H(z) + SNeIa + BAO$, around the best fit point (0.062845, 65.4800) shown in Fig. 3(b).

The likelihood contours for the parameters $\alpha$ and $H_0$ with $1\sigma$, $2\sigma$ and $3\sigma$ errors in the plane $\alpha$-$H_0$ are shown in Figs. 2 and 3. The good fit values of $\alpha$ are constrained 0.216746, 0.007485, 0.145958 and 0.062845 by employing the Hubble dataset $H(z)$, $BAO$, and combined dataset $H(z) + SNeIa$ and $H(z)+SNeIa+BAO$ for which the corresponding good fit values of $H_0$ in the units of $Km/s/Mpc$ are constrained 65.4940, 65.0100, 65.4934 and 65.4800, respectively (see Table 1). Also the good fit value of the model parameter $\alpha = 0.128497$ and Hubble constant $H_0 = 66.4581 \, Km/s/Mpc$ with $1\sigma$, $2\sigma$ and $3\sigma$ errors are constrained in the ranges $0.008477 < \alpha < 0.2474$, $-0.07116 < \alpha < 0.3209$ and $-0.1477 < \alpha < 0.3944$ respectively according to the $SNeIa$ (Union 2.1 compilation data).

## 4 Statistical analysis of cosmological parameters of the model

The present values of various cosmological parameters of our model are computed by the constrained value of the model parameter $\alpha$ using different observational datasets $H(z)$, $BAO$, and combined dataset $H(z) + SNeIa$ and $H(z) + SNeIa + BAO$, respectively (see Table 2).

In the expansion history of the Universe, the two physical parameters $H$ and $q$ are very useful. This can be illustrated as $H > 0$ or $\dot{a} > 0$ specifies an expanding Universe whereas $q < 0$ or $\ddot{a} > 0$ specifies accelerated expanding Universe. In order to discuss accelerated expansion of the Universe, various dark energy models have proposed in the literature. The deceleration parameter $q$ is probed to be very dominant cosmological parameters among other parameters which trace the dynamics of the Universe. Here, we review the different phases of the evolution of the deceleration parameter. The Universe undergoes a cosmic acceleration in late times with a slower rate of expansion, which is revealed by the cosmic observations [1,2]. Moreover, the decelerating phase of the

**Table 2** Summary of the numerical results of cosmological parameters of the model

| Dataset | $q_0$ | $z_{tr}$ | $z_{tr}^*$ | $j_0$ | $\omega_{\phi 0}$ | $\omega_{total 0}$ | $\Omega_{m0}$ | $\Omega_{\phi 0}$ |
|---|---|---|---|---|---|---|---|---|
| $H(z)$ (29 points data) | $-0.477$ | $\simeq 0.841$ | $\simeq -0.866$ | 0.517 | $-0.869$ | $-0.651$ | 0.251 | 0.749 |
| $SNeIa$ (Union 2.1 compilation data) | $-0.540$ | $\simeq 0.828$ | $\simeq -0.940$ | 0.699 | $-0.922$ | $-0.693$ | 0.248 | 0.751 |
| $BAO$ | $-0.626$ | $\simeq 0.819$ | $\simeq -0.984$ | 0.981 | $-0.996$ | $-0.751$ | 0.246 | 0.754 |
| $H(z) + SNeIa$ | $-0.527$ | $\simeq 0.836$ | $\simeq -0.928$ | 0.661 | $-0.912$ | $-0.684$ | 0.249 | 0.750 |
| $H(z) + SNeIa + BAO$ | $-0.586$ | $\simeq 0.827$ | $\simeq -0.966$ | 0.847 | $-0.962$ | $-0.724$ | 0.247 | 0.752 |

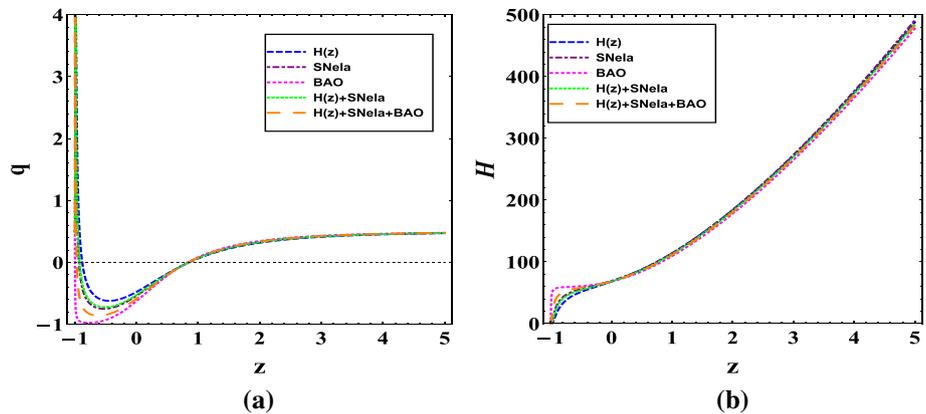

**Fig. 4 a** The plot of deceleration parameter $q$ vs. redshift $z$. **b** The plot of Hubble parameter $H$ vs. redshift $z$ for different observational datasets

Universe is obligatory for the structure formation. The Phase transition of $q$ from deceleration to acceleration is essential to trace the dynamics of the Universe.

In this model, the present accelerating behaviour can be seen by observing the values of the $q$ with respect to redshift $z$ with two transition phases. Taking the model parameter $\alpha$ from the various observational data (see Table 1), we plot the graph of $q$ with respect to redshift $z$ using Eq. (23). The first phase transition ($z_{tr}$) occurs from early deceleration to the acceleration in the quintessence region and the second phase transition ($z_{tr}^*$) occurs form the acceleration in the quintessence region to deceleration and stops the expansion of the Universe in late times (see Fig. 4a, b). In the Fig. 4a, we also observed that $DP$ varies from positive to negative according as the variation of redshift $z$ from high and low, respectively in the range $0.216746 \leq \alpha \leq 0.007485$ of the model parameter $\alpha$ using different observational datasets $H(z)$, $H(z) + SNeIa$, $SNeIa$, $H(z) + SNeIa + BAO$ and $BAO$. The phase transitions occur in the range $0.841 \leq z_{tr} \leq 0.819$. i.e. the Universe goes through a possible phase transition around the near past $z_{tr} \approx 0.8$ from deceleration to acceleration which is consistent with the observations proposed by Amendola [82]. The expansion of the Universe stops suddenly after the second phase transition in late times at $z \simeq -1$ (see Fig. 4a, b). This shows the negative inflation in the Universe.

It is observed that the slope of the $q - z$ curve is minimum and maximum i.e. the Universe shows minimum and maximum acceleration, and the variation of model parameter $\alpha$ maximum and minimum according as $H(z)$ and $BAO$ observational datasets respectively at present. Here, the slope of the $q-z$ curve increases as the variation of model parameter $\alpha$ decreases according as the observational datasets $H(z)$, $H(z) + SNeIa$, $SNeIa$, $H(z) + SNeIa + BAO$, $BAO$, respectively (see Fig. 4a). Therefore, it is noticed that the our model exhibits deceleration at early stages of the evolution of the Universe, acceleration in the quintessence region and starts deceleration after second phase transition until the expansion of the Universe stops in late times. This model shows a quintessence dark energy model at present.

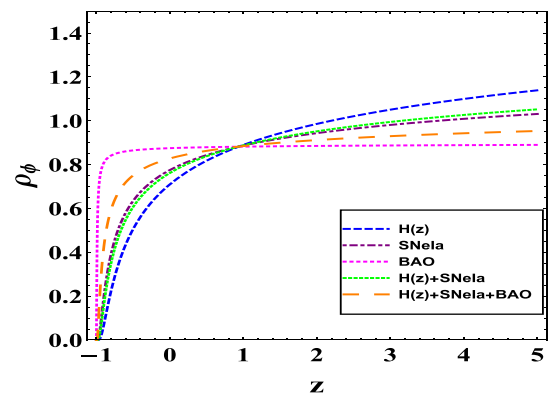

**Fig. 5** The plot of energy density of scalar field $\rho_\phi$ vs. redshift $z$ for different observational datasets

Combining Eqs. (15) and (16), we plot the curves of the energy density of scalar field $\rho_\phi$ with respect to the redshift $z$ for all constraints values of model parameter $\alpha$ obtained from different observational datasets $H(z)$, $SNeIa$, $BAO$, $H(z) + SNeIa$ and $H(z) + SNeIa + BAO$ (see Fig. 5). In this plot, the slope of the $\rho_\phi - z$ curve decreases as the variation of model parameter $\alpha$ decreases according as the observational datasets $H(z)$, $SNeIa$, $BAO$, $H(z)+SNeIa$ and $H(z) + SNeIa + BAO$ respectively. Thus, we see that the energy density of scalar field is high initially at the time of evolution of the Universe and eventually $\rho_\phi \to 0$, which indicates that the amount of dark energy density drops off to nearly zero at late times.

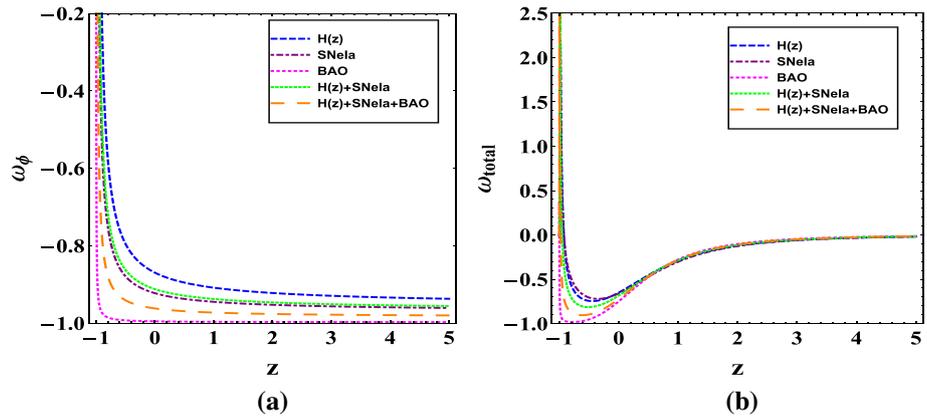

**Fig. 6** **a** and **b** Show the behaviours of EoS parameter of scalar field $\omega_\phi$ and total EoS parameter $\omega_{total}$ w.r.t. redshift $z$

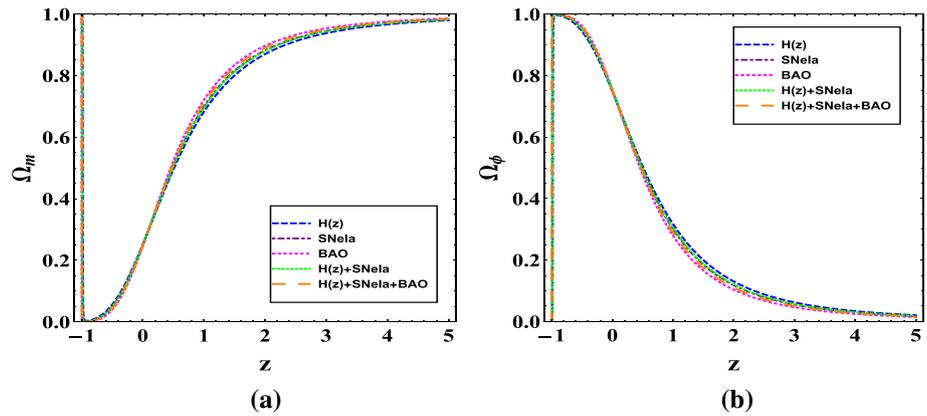

**Fig. 7** **a** and **b** Highlight the behaviours of density parameter of matter and scalar field $(\Omega_m, \Omega_\phi)$ w.r.t. redshift $z$

As it is known that the different stages of the cosmic evolution of the Universe can be analysed by study the behaviour of EoS parameter $\omega$. In our model, we have discussed the EoS parameter of scalar field $\omega_\phi$ and the total EoS parameter $\omega_{total}$. In GR, the only condition to achieve inflation in the universe is $1 + 3\omega < 0$, which produces repulsive energy and jerk in the Universe. The five trajectories of EoS parameter of scalar field $\omega_\phi$ for different values of the model parameter $\alpha$ constrained in different observational datasets $H(z)$, $SNeIa$, $BAO$, $H(z)+SNeIa$ and $H(z)+SNeIa+BAO$ are highlighted in Fig. 6a. The current value of EoS parameter of scalar field $\omega_{\phi_0}$ for different observational datasets given in Table 2 are consistent with the constraint range given by Riess [83]. In Fig. 6a, it is observe that the range $-1 < \omega_\phi < -0.2$ of EoS parameter of scalar field suggests that scalar field $\phi$ starts its evolution from the quintessence region [47].

Figure 6b depicts the profile of total EoS parameter $\omega_{total}$ with respect to redshift $z$. It can be observed that $\omega_{total}$ starts from the quintessence region at the time of evolution of the Universe remains in the same region at present time and suddenly collapsed at late times.

Figure 7a, b demonstrate the contrasting behaviour of density parameters of matter field $\Omega_m$ and scalar field $\Omega_\phi$ w.r.t. redshift $z$. It can be noticed that $\Omega_m$ is dominated over $\Omega_\phi$ at the early phases of the evolution of the Universe. The density parameters of matter field $\Omega_m$ is monotonically decreasing while the density parameters of scalar field is monotonically increasing in late times. Moreover, both parameters $\Omega_m$ and $\Omega_\phi$ suddenly collapsed as $z \longrightarrow -1$ which shows the big rip singularity in infinite future. Moreover, the sum of these two parameters is equal to unity, which is consistent with different observational data.

The history of the expanding universe can be interpreted in a better way by analysing the behaviour of parameters, which involves some higher derivatives of the scale factor. The geometrical behaviour of these parameters provides a measure to examine the performance of various dark energy models. Here, we intend to discuss a kinematic quantity jerk parameter $j$, which is obtained by the third order derivative of the scale factor and is defined as [84,85]

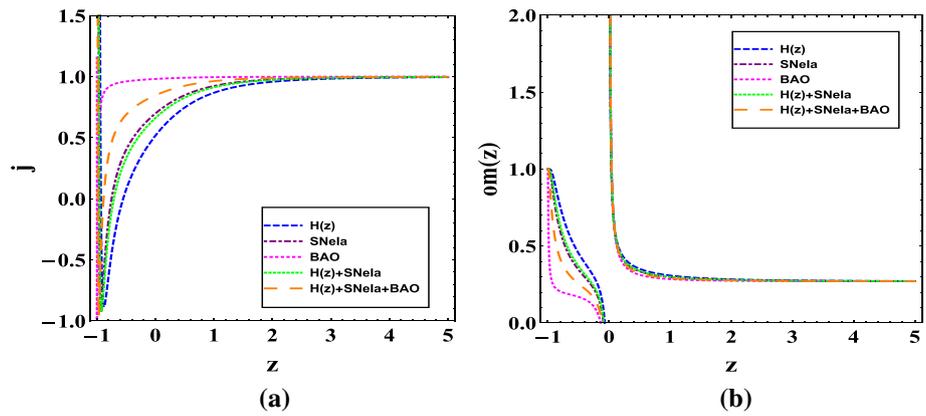

**Fig. 8** **a** and **b** Represent the jerk parameter of our model and om diagnostic w.r.t. redshift z

**Table 3** Summary of the numerical results of cosmographic parameters of the model

| Dataset | $j_0$ | $s_0$ | $l_0$ | $m_0$ |
| --- | --- | --- | --- | --- |
| $H(z)$ (29 points data) | 0.5174 | −0.9042 | 2.0671 | −9.5617 |
| $SNeIa$ (Union 2.1 compilation data) | 0.6993 | −0.6674 | 2.0018 | −9.1940 |
| $BAO$ | 0.9813 | −0.1467 | 2.4938 | −7.6850 |
| $H(z) + SNeIa$ | 0.6617 | −0.7233 | 1.9919 | −9.2889 |
| $H(z) + SNeIa + BAO$ | 0.8476 | −0.4149 | 2.1664 | −8.6155 |

$$j = \frac{\dddot{a}}{aH^3}. \quad (38)$$

According to the standard $\Lambda CDM$ model, the value of jerk parameter is equal to unity i.e. $j = 1$. The deviation from $j = 1$ investigates the dynamics of different kinds of dark energy models other than standard model $\Lambda CDM$. Figure 8a highlights the evolution of jerk parameter w.r.t. redshift z for the parametrization of energy density of scalar field $\rho_\phi$ (13). It has been observed that our model behaves like standard $\Lambda CDM$ model at the time of evolution of the early Universe and represents the other DE model different from $\Lambda CDM$ at present time as $j$ is different from 1 for all observational dataset (see Table 2).

The higher order cosmographic parameters in terms of redshift z can be expressed as [86]

$$\begin{aligned}
j &= -q + 2q(1+q) + (1+z)\frac{dq}{dz}, \\
s &= j - 3j(1+q) - (1+z)\frac{dj}{dz}, \\
l &= s - 4s(1+q) - (1+z)\frac{ds}{dz}, \\
m &= l - 5l(1+q) - (1+z)\frac{dl}{dz}.
\end{aligned} \quad (39)$$

The present value of these cosmographic parameters $j_0$, $s_0$, $l_0$ and $m_0$ can be obtained by using various observational datasets, which is shown in Table 3.

Apart from the several techniques which have been used to measure the contrast of $\Lambda CDM$ model among various DE models is a diagnostic tool known as Om diagnostic [87,88]. This technique is used to distinguish various DE models without comprising the density parameter of matter and EoS Parameter. It investigates the behaviour of different trajectories of $Om(z)$ with respect to redshift z and is evaluated by a single parameter $H$ with redshift z. It is formulated as

$$Om(z) = \frac{\{\frac{H(z)}{H_0}\}^2 - 1}{z^3 + 3z^2 + 3z}. \quad (40)$$

The slope of the function $Om(z)$ decides the behaviour of DE model. The model represents quintessence DE model when the curvature of the $Om(z)$ w.r.t. z is negative and the model represents a phantom DE model when the curvature of $Om(z)$ w.r.t. z is positive. Also model is similar to $\Lambda CDM$ when $Om(z)$ has zero curvature. In Fig. 8b, it can be seen that all the trajectories of $Om(z)$ increase as redshift z decreases. Hence, the negative curvature pattern represents that our model is a quintessence DE model for all the observational dataset. Moreover, In Fig. 8b, the gap between the two sets of the trajectories indicates the presence of dark matter, which may be expected in the hallow region of quintessence dark energy.

## 5 Energy conditions

In General Relativity, there are various prevailing energy conditions (ECs) which impose certain conditions to prevent such regions having negative energy density. In other words, ECs are treated as a viable generalization of the fact that energy density in the Universe can never be negative to the whole EMT [89–91]. There are wide applications of ECs to examine the viability of many important singularity problems of space time, black holes and wormholes etc. Here, many ECs are commonly used in GR whose viabilities can be studied by the famous Raychaudhuri equation [92]. The energy conditions can be mostly stated in two ways: (i) geometric way, where ECs are well expressed in terms of Ricci tensor or Weyl tensor, and (ii) physical way, where ECs are expressed in stress energy momentum tensor or expressed in terms of energy density $\rho$, which performs as the time-like component and pressures $p_i$, $i = 1, 2, 3$, which represent the 3-space-like component). The standard point-wise ECs in terms of $\rho$ and $p$ in GR are defined as:

- Null energy condition (NEC) $\Leftrightarrow \rho + p_i \geq 0, \forall i$,
- Weak energy condition (WEC) $\Leftrightarrow \rho \geq 0, \rho + p_i \geq 0, \forall i$,
- Strong energy condition (SEC) $\Leftrightarrow \rho + \sum_{i=1}^{3} p_i \geq 0, \rho + p_i \geq 0, \forall i$,
- Dominant energy condition (DEC) $\Leftrightarrow \rho \geq 0, \rho \geq |p_i|, \forall i$, where $i = 1, 2, 3$.

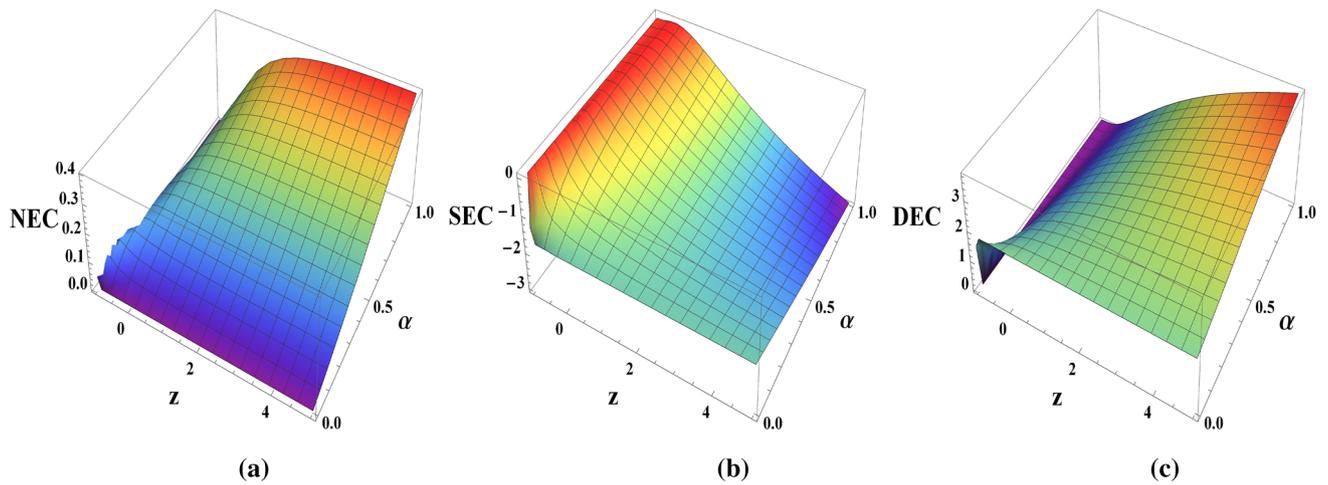

**Fig. 9** The plots of energy conditions NEC, SEC and DEC for the model

Also we know that the energy density $\rho$ and isotropic pressure $p$ can be described in the form of potental energy $V(\phi)$ and kinetic energy $\frac{\dot\phi^2}{2}$ of the model i.e. in terms of scalar field $\phi$. Thus, the standard point-wise ECs in GR in terms of scalar field $\phi$ are defined as:

- NEC: $\forall V(\phi)$
- WEC $\Leftrightarrow V(\phi) \geq \frac{\dot\phi^2}{2}$,
- SEC $\Leftrightarrow V(\phi) \leq \dot\phi^2$.
- DEC $\Leftrightarrow V(\phi) \geq 0$.

From Fig. 9, we observe that NEC, WEC and DEC are satisfied but SEC is violating in our model, which accomplishes that our model exhibits accelerating expansion in the Universe. The fulfilment of NEC, WEC and DEC in this model is somewhat the mandatory requirement in the Universe to impose certain conditions which prevent from such macroscopic regions having negative energy density. However, the violation of SEC indicates the presence of some exotic matter in the Universe to produce anti gravitational effects (see Fig. 9b). This shows that inflation occurs because $\rho_\phi + 3p_\phi < 0$, which implies that inflation occurs due to $V(\phi) > \frac{\dot\phi^2}{2}$ [51]. Hence the Universe is likely to enter into possible reheating stage after inflation and the expansion of the Universe is accelerating under the influence of a repulsive force which appears to be a strong enough thrust.

## 6 Dynamics of reheating after inflation

In order to explain the accelerating universe, we need $\ddot{a} > 0$, the Friedmann Eqs. (2) and (3) requires $\rho_\phi + 3p_\phi < 0$ This implies that inflation occurs as long as $\dot\phi^2 < V(\phi)$. The Eqs. (2), (3) and the Euler Lagrange equation for $\phi$ following from the action in (4) and (5) are

$$H^2 = \frac{8\pi G}{3}\left[\frac{\dot\phi^2}{2} + V(\phi)\right], \tag{41}$$

and the Klein Gordon equation for the scalar field

$$\ddot\phi + 3H\dot\phi + \partial_\phi V(\phi) = 0. \tag{42}$$

The most remarkable approach of particle creation in the quantum theory is the Reheating Theory of the Universe after inflation since almost all matter constituents of the Universe were created at the subsequent radiation dominated era during this process [65]. It is noticeable that the total energy was accumulated in a classical slowly moving inflaton field $\phi$ at this stage of inflation and this field $\phi$ started oscillat-ing with its minimum effective potential. Steadily, it created the elementary particles, they interacted with each other and reached to a thermal equilibrium state with a temperature $T_R$, which was termed as the reheating temperature.

Again in a simple chaotic inflation scenario which explains the classical inflaton scalar field $\phi$, we consider the effective potential $V(\phi) = \pm\frac{1}{2}m_\phi{}^2\phi^2 + \frac{\lambda}{4}\phi^4$, where $m_\phi$ is the mass of scalar field $\phi$. The term $\frac{1}{2}m_\phi{}^2\phi^2$ is assumed as the inflaton potential and the negative sign indicates the spontaneous symmetry breaking $\phi \longrightarrow \phi + \sigma$ with creation of $\sigma = \frac{m_\phi}{\sqrt{\lambda}}$ called as a classical scalar field. After inflation, the classical inflaton scalar field $\phi$ may decay into two very small particles bosons $\chi$ and fermions $\psi$ due to the interaction terms $-\frac{1}{2}g^2\phi^2\chi^2$ and $-h\bar{\psi}\psi\phi$, where $\lambda$, $g$ and $h$ are taken negligibly small coupling constants. Since the masses of the particles $\chi$ and $\psi$ are negligibly small, therefore we can assume $m_\chi\phi = g\phi$, $m_\psi\phi = |h\phi|$ [65].

In a study of most basic reheating theory [69], we have a inflation phase at $\phi > M_P$ which is favoured by the friction like term $3H\dot{\phi}$ in Eq. (42). As the inflaton scalar field $\phi$ decreases the effect of the term $3H\dot{\phi}$ decreases, the inflation vanishes at $\phi < \frac{1}{2}M_P$ and the field $\phi$ starts oscillating in the vicinity of $V(\phi)_{min}$ [65]. The amplitude of the oscillations slowly decreases due to expansion of the universe and the energy transfer to particles created by the oscillating field. The assumption of the classical oscillating scalar field $\phi(t)$ indicated as a cluster of scalar particles at rest explains the basic reheating theory. Then the decay rate in the energy of oscillations may be equated to the decay rate of $\phi$-particles. Therefore, the decay rates are given by

$$\Gamma_{(\phi \longrightarrow \chi\chi)} = \frac{g^4\sigma^2}{8\pi m_\phi}, \quad \Gamma_{(\phi \longrightarrow \psi\psi)} = \frac{h^2 m_\phi}{8\pi}. \tag{43}$$

The reheating of the Universe accomplishes when the rate of expansion of the universe $H$ becomes smaller than the total decay rate in the energy of oscillations $\Gamma_{total}$, where $\Gamma_{total} = \Gamma_{(\phi \longrightarrow \chi\chi)} + \Gamma_{(\phi \longrightarrow \psi\psi)}$ and $H = \sqrt{\frac{8\pi m_\phi}{3M_P^2}}$ which is equivalent to $t^{-1}$. The reheating temperature may be approximated by $T_R = 0.1\sqrt{\Gamma_{total}M_P}$ [65].

With this analytical view, we can mention that the Reheating Theory yields a conditionally exact interpretation of particle decay at the last phases of reheating. Moreover, this theory is consistently relevant when the inflaton field decays into fermions only with a negligibly small coupling constant $h^2 \ll \frac{m_\phi}{M_p}$. However, this theory fails to interpret the first stages of reheating. In order to establish the theory of the first stage of reheating, we start this theory with massive scalar field $\phi$ which decays into $\chi$-particles, then we review $\frac{\lambda}{4}\phi^4$ theory, and finally we discuss reheating at the first stage with spontaneous symmetry breaking theories [65]. Therefore, we can observe that with this diagnostic tool, we may be able to acquire the appropriate information of physics in the early universe to discuss the reheating phase after inflation in our present EDSFD parametric model.

## 7 Discussions and conclusions

In this work, we have presented an FLRW cosmological model by taking an appropriate parametrization in a specific differential equation in terms of energy density of scalar field $\rho_\phi$, which causes a phase transition from early deceleration to present cosmic acceleration. Using this parametrization, we have reconstructed the equation of state parameter $\omega_\phi(z)$ to discuss the evolution of the early universe in a flat FLRW space time. We have constrained the model parameter using the various observational datasets of Hubble parameter $H(z)$, latest Union 2.1 compilation dataset $SNeIa$, $BAO$, joint dataset $H(z) + SNeIa$ and $H(z) + SNeIa + BAO$, and we have also fixed its best fit value at present.

This article is an attempt to construct a cosmological model by assuming a specific form of a cosmic evolution scenario. As we have already studied in details the advantages of parametric reconstruction method, therefore to understand the hidden realities of the cosmos, this method assists one to construct a cosmological model which can predict the possible phase transition. So in somewhat, the idea of the parametric approach is quite reasonable, simple and help to improve the efficiency of future cosmological scenario and open the doors to understand the nature of dark energy. The particular and quite arbitrary form of a function of scale factor $a$ has been chosen in such a manner which contribute a genuine form of the energy density of scalar field $\rho_\phi$, having a decreasing slope at each point of time. The motivation for taking the parametrization of $\rho_\phi$ has been discussed in the introduction section [49,50]. Our main intention in this paper is to consider a quintessence like scalar field as a candidate of dark energy for studying the dynamics of present accelerating universe in the framework of flat FLRW space time. In accordance with, we have recorded the main outcomes of our investigation of the model as follows:

- In Sect. 3, the error bar plots of Hubble dataset and $SNeIa$ Union 2.1 compilation dataset represent that both panels are fitted well while correlating our model with $\Lambda CDM$ using observational dataset $H(z)$ and $SNeIa$ Union 2.1 compilation data respectively (see Fig. 1a, b).
- Figures 2 and 3 depicts the likelihood contours for the parameters $\alpha$ and $H_0$ with $1\sigma$, $2\sigma$ and $3\sigma$ errors in the plane $\alpha$-$H_0$. The good fit values of $\alpha$ are constrained 0.216746, 0.007485, 0.145958 and 0.062845 by employing the Hubble dataset $H(z)$, $BAO$, combined dataset $H(z) + SNeIa$ and $H(z) + SNeIa + BAO$ for which the corresponding good fit values of $H_0$ in the units of $Km/s/Mpc$ are constrained 65.4940, 65.0100, 65.4934 and 65.4800 respectively (see Table 1). Also the good fit value of the model parameter $\alpha = 0.128497$ and Hubble constant $H_0 = 66.4581\ Km/s/Mpc$ with $1\sigma$, $2\sigma$ and $3\sigma$ errors are constrained in the ranges $0.008477 < \alpha < 0.2474$, $-0.07116 < \alpha < 0.3209$ and $-0.1477 < \alpha < 0.3944$, respectively according to the $SNeIa$ (Union 2.1 compilation data).

- In Sect. 4, it is noticed that the our model exhibits deceleration at early stages of the evolution of the Universe, acceleration in the quintessence region and starts deceleration after second phase transition until the expansion of the Universe stops in late times. The early decelerating phase of the Universe is capable to describe the structure formation in the Universe. This model shows a quintessence dark energy model at present (see Fig. 4a, b).
- Figure 5 explains that the assumed parametrization of energy density of scalar field is high initially at the time of evolution of the Universe and eventually $\rho_\phi \rightarrow 0$, which indicates that the amount of dark energy density diminishes to zero at late times.
- In Fig. 6a, it is observed that the range $-1 < \omega_\phi < -0.2$ of EoS parameter of scalar field suggests that scalar field $\phi$ starts its evolution from the quintessence region [47]. Figure 6b depicts that $\omega_{total}$ starts from the quintessence region at the time of evolution of the Universe remains in the same region at present time and suddenly collapsed in late times.
- Figure 7a, b demonstrate the contrasting behaviour of density parameters of matter field $\Omega_m$ and scalar field $\Omega_\phi$ w.r.t. redshift $z$. It can be noticed that $\Omega_m$ is dominated over $\Omega_\phi$ at the early phases of the evolution of the Universe. The density parameters of matter field $\Omega_m$ is monotonically decreasing while the density parameters of scalar field is monotonically increasing in late times. Moreover, Both parameters $\Omega_m$ and $\Omega_\phi$ suddenly collapsed as $z \longrightarrow -1$ which shows the big rip singularity in infinite future. Moreover, the sum of these two parameters is equal to unity, which is consistent with different observational data.
- In Fig. 8a, it is concluded that our model behaves like standard $\Lambda CDM$ model at the time of evolution of the early Universe and represents the other DE model different from $\Lambda CDM$ at present time since $j$ is different from 1 for all observational dataset (see Table 2). Figure 8b enacts that all the trajectories of $Om(z)$ increase as redshift $z$ decreases. Hence, the negative curvature pattern represents that our model is a quintessence DE model for all the observational dataset. Moreover, in this figure the gap between the two sets of the trajectories indicates the presence of dark matter, which may be expected in the hallow region of quintessence dark energy.
- In Fig. 9, we observe that NEC and DEC are satisfied but SEC is violating in our model, which accomplishes that our model exhibits accelerating expansion in the Universe. The NEC and DEC in this model is satisfied, which is somewhat the mandatory requirement in the Universe to impose certain conditions which prevent from such macroscopic regions having negative energy density. However, the violation of SEC is the same as the presence of some exotic matter in the Universe to produce anti gravitational effects. Hence the Universe is likely to enter into reheating stage after inflation and the expansion of the Universe is accelerating under the influence of a repulsive force which appears to be a strong enough thrust.
- In Sect. 6, it is analytically mentioned that the Reheating Theory of the Universe is developed after inflation. We observe that the first stage of reheating takes place when the classical inflaton field $\phi$ decays either into $\phi$-particles or into other bosons abruptly due to a parametric resonance on a large scale. Subsequently these bosons decay into some other particles, which become thermalized in long run. The complete reheating is possible only when a single $\phi$-particle decays into other particles. This inflicts best fit constraints on the structure of the inflationary models and indicates that the inflaton field may be a dark matter candidate [65]. Therefore, we can observe that with this diagnostic tool, we may be able to acquire the appropriate information of physics in the early universe to discuss the reheating stage after inflation in our present EDSFD parametric model.

Here, after examining the above points, we can conclude that our present cosmological model is found to be very interesting with the approach of reconstruction of some physical parameters by EDSFD parametrization, and contribute a reasonable domain of knowledge to understand the various cosmological scenario of the evolution of the early Universe. Undoubtedly, the inclusion of some observational datasets in this study provides a more precise range to model parameters so that the behaviour of geometrical and physical parameters could be investigated more appropriately. However, the present article is only an attempt in the direction to study the nature of dark energy. Moreover, the advancement of the observational cosmology in our model will provide one suitable step to upgrade our understanding and make us better to interpret the mysterious nature of dark energy. In our present EDSFD parametric model, the dynamics of reheating stage after inflation can be discussed. The complete reheating inflicts the best fit constraints on the structure of the inflationary models which indicates that the inflaton field may be a dark matter candidate.

**Acknowledgements** The author J. K. Singh expresses his thanks to Prof. M. Sami and Prof. S. G. Ghosh, CTP, Jamia Millia Islamia, New Delhi, India for some fruitful discussions. Authors also express their thanks to the referee for his valuable comments and suggestions.

**Data Availability Statement** This manuscript has no associated data or the data will not be deposited. [Authors' comment: (i) My work is a theoretical work and no experimental data has been used, (ii) The data used during the study are contained in this article.]

## References

1. A.G. Riess et al., Astron. J. **116**, 1009 (1998)
2. S. Perlmutter et al., Astrophys. J. **517**, 565 (1999)
3. M. Tegmark et al., Phys. Rev. D **69**, 103501 (2004)
4. U. Seljak et al., Phys. Rev. D **71**, 103515 (2005)
5. G. Hinshaw et al., Astrophys. J. Suppl. **208**, 19 (2013)
6. C.L. Bennett et al., Astrophys. J. Suppl. **208**, 20 (2013)
7. L. Anderson et al., Mon. Not. R. Astron. Soc. **427**, 3435 (2013)
8. D.J. Eisenstein et al., Astrophys. J. **633**, 560 (2005)
9. B. Jain, A. Taylor, Phys. Rev. Lett. **91**, 141302 (2003)
10. Y. Akramiet et al., (Planck 2018 results). arXiv:1807.06205v1
11. N. Aghanim et al., (Planck 2018 results) arXiv:1807.06209v1
12. S. Rani, A. Altaibayeva, M. Shahalam, J.K. Singh, R. Myrzakulov, J. Cosm. Astropart. Phys. **03**, 031 (2015)
13. J.K. Singh, S. Rani, Appl. Math. Comp. **259**, 187 (2015)
14. J.K. Singh, N.K. Sharma, A. Beesham, Appl. Math. Comp. **270**, 567 (2015)
15. I. Leanizbarrutia, Francisco S.N. Lobo, D. Saez-Gomez, Phys. Rev. D **95**, 084046 (2015)
16. S. Weinberg, Rev. Mod. Phys. **61**, 1 (1989)
17. P.J. Steinhardt, L.M. Wang, I. Zlatev, Phys. Rev. D **59**, 123504 (1999)
18. B.D. Normann, I. Brevik, Mod. Phys. Lett. A **32**, 1750026 (2017)
19. I. Brevik, Ø. Grøn, J. de Haro, S.D. Odintsov, E.N. Saridakis, Int. J. Mod. Phys. D **26**, 1730024 (2017)
20. P.J. Peebles, R. Ratra, Astrophys J. **325**, L17 (1988)
21. C. Wetterich, Astron. Astrophys. **301**, 321–328 (1995)
22. R.R. Caldwell, Rahul Dave, Paul J. Steinhardt, Phys. Rev. Lett. **80**, 1582 (1998)
23. M. Sami, T. Padmanabhan, Phys. Rev. D **67**, 083509 (2003)
24. J. Khoury, A. Weltman, Phys. Rev. Lett. **93**, 171104 (2004)
25. J.K. Singh, N.K. Sharma, A. Beesham, Theor. Math. Phys. **193**, 1865 (2017)
26. E.J. Copeland, M. Sami, S. Tsujikawa, Int. J. Mod. Phys. D **15**, 1753–1936 (2006)
27. J. Martin, Mod. Phys. Lett. A **23**, 1252–1265 (2008)
28. W. Zimdahl, D. Pavn, L.P. Chimento, Phys. Lett. B **521**, 133–138 (2001)
29. L. Amendola, Phys. Rev. D **62**, 043511 (2000)
30. M.S. Turner, A.G. Riess, Astrophys. J. **569**, 18 (2002)
31. G.F.R. Ellis, M.S. Madsen, Class. Quant. Grav. **8**, 667 (1991)
32. A.A. Starobinsky, J. Exp. Theor. Phys. Lett. **68**, 757 (1998)
33. Y. Gong, A. Wang, Phys. Rev. D **75**, 043520 (2007)
34. B. Santos et al., Astropart. Phys. **35**(2011), 17 (2011)
35. R. Nair et al., J. Cosm. Astropart. Phys. **01**, 018 (2012)
36. A. Shafieloo, Mon. Not. R. Astron. Soc. **380**, 1573 (2007)
37. R.G. Crittenden et al., J. Cosm. Astropart. Phys. **02**, 048 (2012)
38. T. Holsclaw et al., Phys. Rev. D **82**, 103502 (2010)
39. J.V. Cunha, J.A.S. Lima, Mon. Not. R. Astron. Soc. **390**, 210 (2008)
40. O. Akarsu et al., Eur. Phys. J. Plus **129**, 22 (2014)
41. L. Xu, J. Lu, Mod. Phys. Lett. A **24**, 369 (2009)
42. J.K. Singh, K. Bamba, R. Nagpal, S.K.J. Pacif, Phys. Rev. D **97**, 123536 (2018)
43. J.K. Singh, R. Nagpal, S.K.J. Pacif, Int. J. Geom. Meth. Mod. Phys. **15**, 1850049 (2018)
44. R. Nagpal, J.K. Singh, S. Aygün, Astrophys. Space Sci. **363**, 114 (2018)
45. R. Nagpal, S.K.J. Pacif, J.K. Singh, Kazuharu Bamba, A. Beesham, Eur. Phys. J. C **78**, 946 (2018)
46. E.V. Linder, Phys. Rev. Lett. **90**, 091301 (2003)
47. R. Nagpal, J.K. Singh, A. Beesham, H. Shabani, Ann. Phys. **405**, 234 (2019)
48. S. Nojiri, S.D. Odintsov, V.K. Oikonomou, Phys. Lett. B **747**, 310 (2015)
49. Y. Wang, P.M. Garnavich, Astrophys. J. **552**, 445 (2001)
50. I. Maor, R. Brustein, J. McMahon, P.J. Steinhardt, Phys. Rev. D **65**, 123003 (2002)
51. K. Lozanov, Lectures on Reheating after Inflation (2019). arXiv:1907.04402
52. R. Micha, I.I. Tkachev, Phys. Rev. Lett. **90**, 121301 (2003)
53. R. Micha, I.I. Tkachev, Phys. Rev. D **70**, 043538 (2004)
54. I. Tkachev, S. Khlebnikov, L. Kofman, A.D. Linde, Phys. Lett. B **440**, 262 (1998)
55. A. Rajantie, E.J. Copeland, Phys. Rev. Lett. **85**, 916 (2000)
56. J.-F. Dufaux, D.G. Figueroa, J. Garcia-Bellido, Phys. Rev. D **82**, 083518 (2010)
57. D.G. Figueroa, J. Garca-Bellido, F. Torrent, Phys. Rev. D **93**, 103521 (2016)
58. P. Adshead, J.T. Giblin, T.R. Scully, E.I. Sfakianakis, J. Cosm. Astropart. Phys. **1512**, 034 (2015)
59. K.D. Lozanov, M.A. Amin, J. Cosm. Astropart. Phys. **1606**, 032 (2016)
60. K.D. Lozanov, M.A. Amin, Phys. Rev. Lett. **119**, 061301 (2017)
61. M.P. Hertzberg, J. Karouby, W.G. Spitzer, J.C. Becerra, L. Li, Phys. Rev. D **90**, 123528 (2014)
62. L. Dai, M. Kamionkowski, J. Wang, Phys. Rev. Lett. **113**, 041302 (2014)
63. J. Martin, C. Ringeval, V. Vennin, Phys. Rev. Lett. **114**, 081303 (2015)
64. J. Martin, C. Ringeval, V. Vennin, Phys. Rev. D **93**, 103532 (2016)
65. L. Kofman, A.D. Linde, A.A. Starobinsky, Phys. Rev. Lett. **73**, 3195 (1994)
66. L. A. Kofman, The Origin of matter in the universe: Reheating after inflation, 1996. arXiv:astro-ph/9605155 [astro-ph]. Accessed 24 May 1996
67. L. Kofman, A.D. Linde, A.A. Starobinsky, Phys. Rev. D **56**, 3258 (1997)
68. B.A. Bassett, S. Tsujikawa, D. Wands, Rev. Mod. Phys. **78**, 537 (2006)
69. Planck Collaboration, P. A. R. Ade et al., Astron. Astrophys. **594** A 20 (2016)
70. J.D. Barrow, Nucl. Phys. B **310**, 743 (1988)
71. J.D. Barrow, Phys. Lett. B **235**, 40 (1990)
72. S. Weinberg, *Cosmology and gravitation* (Wiley, New York, 1972)

73. N. Suzuki et al., Astrophys. J. **746**, 85 (2012)
74. N. Padmanabhan, X. Xu, D.J. Eisenstein, R. Scalzo, A.J. Cuesta, K.T. Mehta et al., Mon. Not. R. Astron. Soc. **427**, 2132 (2012)
75. F. Beutler, C. Blake, M. Colless, D.H. Jones, L. Staveley-Smith, L. Campbell et al., Mon. Not. R. Astron. Soc. **416**, 3017 (2011)
76. BOSS collaboration, L. Anderson et al., Mon. Not. R. Astron. Soc., **441** 24 (2014)
77. C. Blake et al., Mon. Not. R. Astron. Soc. **425**, 405 (2012)
78. P. A. R. Ade et al., Planck 2015 results. XIII. Cosmological parameters, Preprint arXiv:1502.01589 (2015)
79. M. Vargas dos Santos, Ribamar R.R. Reis, J. Cosm. Astropart. Phys. **1602**, 066 (2016)
80. R. Giostri, M.V.d Santos, I. Waga, R.R.R. Reis, M.O. Calvao, B.L. Lago, J. Cosm. Astropart. Phys. **1203**, 027 (2012)
81. G. Hinshaw et al., Astrophys. J. Suppl. **208**, 19 (2013)
82. L. Amendola, Mon. Not. R. Astron. Soc. **342**, 221 (2003)
83. A.G. Riess, Astrophys. J. **607**, 665 (2004)
84. V. Sahni, T.D. Saini, A.A. Starobinsky, U. Alam, JETP Lett. **77**, 201 (2003)
85. U. Alam, V. Sahni, T.D. Saini, A.A. Starobinsky, Mon. Not. R. Astron. Soc. **344**, 1057 (2003)
86. Yu. L. Bolotin, V. A. Cherkaskiy, O. Yu. Ivashtenko, M. I. Konchatnyi, L. G. Zazunov. arXiv:1812.02394v1
87. V. Sahni, A. Shafieloo, A.A. Starobinsky, Phys. Rev. D **78**, 103502 (2008)
88. C. Zunckel, C. Clarkson, Phys. Rev. Lett. **101**, 181301 (2008)
89. J. Santos, J.S. Alcaniz, Phys. Lett. B **619**, 11 (2005)
90. J. Santos et al., Phys. Rev. D **76**, 043519 (2007)
91. A.A. Sen, R.J. Scherrer, Phys. Lett. B **659**, 457 (2008)
92. S.M. Carroll, *Spacetime and geometry: an introduction to general relativity* (Addison Wesley, Boston, 2004)